\PassOptionsToPackage{usenames,svgnames}{xcolor}

\documentclass[a4, 11pt, parskip=half, DIV=10]{scrartcl}
\usepackage[english]{babel}
\usepackage[T1]{fontenc}
\usepackage[utf8]{inputenc}

\usepackage{graphicx}
\usepackage[numbers]{natbib}
\usepackage{subcaption}
\usepackage{booktabs}
\usepackage{url}
\usepackage{todonotes}
\usepackage{multirow}
\usepackage{csquotes}
\usepackage{paralist}

\usepackage{pgfplots}
\usepgfplotslibrary{colorbrewer}

\pgfdeclarelayer{background}
\pgfsetlayers{background,main}

\usepackage{bm}

\usepackage{amsmath}
\usepackage{amssymb}
\usepackage{amsthm}
\usepackage{amsfonts}
\usepackage{thmtools}		
\usepackage{mleftright}
\usepackage{stmaryrd}
\usepackage{nicefrac}

\theoremstyle{definition}
\newtheorem{theorem}{Theorem}

\newtheorem{definition}[theorem]{Definition}

\newtheorem{remark}[theorem]{Remark}

\usepackage{thm-restate}
\usepackage[mathic=true]{mathtools}
\usepackage{fixmath}
\usepackage{siunitx}
\usepackage{color}

\usepackage{pifont}

\usepackage{enumitem}
\setlist[enumerate]{itemsep=0.2ex, topsep=0.5\topsep}
\setlist[description]{itemsep=0.2ex, topsep=0.5\topsep}
\setlist[itemize]{itemsep=0.2ex, topsep=0.5\topsep}

\usepackage{setspace}
\usepackage{ellipsis}
\usepackage{xspace}
\usepackage{hfoldsty}

\usepackage{abstract}
\usepackage{tcolorbox}
\usepackage{lmodern}
\usepackage[tt=false]{libertine}
\usepackage[varqu]{zi4}
\usepackage[libertine]{newtxmath}
\usepackage[
pdfa,
hidelinks,
pdftex, 
pdfdisplaydoctitle,
pdfpagelabels,
pdfauthor={},
pdftitle={},
pdfsubject={},
pdfkeywords={},
pdfproducer={Latex with the hyperref package},
pdfcreator={pdflatex}
]{hyperref}

\makeatletter
\def\thmt@refnamewithcomma #1#2#3,#4,#5\@nil{%
	\@xa\def\csname\thmt@envname #1utorefname\endcsname{#3}%
	\ifcsname #2refname\endcsname
	\csname #2refname\expandafter\endcsname\expandafter{\thmt@envname}{#3}{#4}%
	\fi
}
\makeatother
\usepackage[capitalise,noabbrev]{cleveref}

\newcommand{\answer}[1]{\paragraph{#1}}
\newcommand{\DB}{\ensuremath{\text{DB}}}
\newcommand{\kNNs}{$k$NN }
\newcommand{\kNN}[2]{\ensuremath{\operatorname{NN}(#1, #2)}}
\newcommand{\range}[2]{\ensuremath{\operatorname{range}(#1, #2)}}
\newcommand{\cNP}[0]{\ensuremath{\mathsf{NP}}}

\usepackage{lipsum}

\usepackage[auth-lg]{authblk}

\recalctypearea

\begin{document}

\title{Metric Indexing for Graph Similarity Search\thanks{This work has been supported by the Vienna Science and Technology Fund (WWTF) through project VRG19-009,
	and Deutsche Forschungsgemeinschaft (DFG), project number 124020371, within the Collaborative Research Center
	SFB 876 ``Providing Information by Resource-Constrained Analysis'', SFB projects A2 and A6.}}
\author[1]{Franka~Bause}
\author[2]{David~B.~Blumenthal}
\author[3]{Erich~Schubert}
\author[1]{Nils M.~Kriege}

\affil[1]{University of Vienna, Faculty of Computer Science, Vienna, Austria }
\affil[ ]{\ttfamily\{franka.bause,nils.kriege\}@univie.ac.at}
\affil[2]{Friedrich-Alexander University Erlangen-Nürnberg (FAU), Department Artificial Intelligence in Biomedical Engineering (AIBE), Erlangen, Germany}
\affil[ ]{\ttfamily david.b.blumenthal@fau.de}
\affil[3]{TU Dortmund University, Department of Computer Science, Dortmund, Germany}
\affil[ ]{\ttfamily erich.schubert@tu-dortmund.de}

\date{\vspace{-50pt}}

\maketitle

\begin{abstract}
Finding the graphs that are most similar to a query graph in a large database is a common task with various applications. A widely-used similarity measure is the graph edit distance, which provides an intuitive notion of similarity and naturally supports graphs with vertex and edge attributes. Since its computation is \cNP-hard, techniques for accelerating similarity search have been studied extensively. However, index-based approaches for this are almost exclusively designed for graphs with categorical vertex and edge labels and uniform edit costs.  
We propose a filter-verification framework for similarity search, which supports non-uniform edit costs for graphs with arbitrary attributes. 
We employ an expensive lower bound obtained by solving an optimal assignment problem. This filter distance satisfies the triangle inequality, making it suitable for acceleration by metric indexing. In subsequent stages, assignment-based upper bounds are used to avoid further exact distance computations.
Our extensive experimental evaluation shows that a significant runtime advantage over both a linear scan and state-of-the-art methods is achieved.  
\end{abstract}

\section{Introduction}
\label{sec:introduction}

Graph-structured data is ubiquitous in many areas such as chemo- and bioinformatics or computer vision. A common task is to search a database containing a large number of graphs for those that are most similar to a given query graph. Such queries are submitted directly by the user or occur as subproblems in downstream machine learning algorithms. 
A widely accepted concept of graph similarity is the \emph{graph edit distance}, which is the minimum cost for transforming one graph into the other by a sequence of edit operations. A strength of this measure is that it can elegantly be applied to graphs with vertex and edge attributes by defining the costs of edit operations adequately.
For example, to compare protein graphs where vertices are annotated by the amino acid sequence of the secondary structure elements they represent, the Levenshtein distance was used~\cite{27_Riesen}.

However, the vast majority of efficient methods for similarity search in graph databases are limited to the special case where graphs have categorical labels and the costs of edit operations are uniform (either zero or one)~\cite{2_ComparingStars,32_GSimJoin,31_SEGOS,12_efficient,3_partitionSim,ZhengZLWZ15,5_simMultiIndex,21_inves2019,CHEN2019762}.
A fairly recent development in this domain are neural graph embeddings, e.g.~\cite{QinBS20}, which do not return exact similarity search results.
For the pairwise computation of the graph edit distance, several exact approaches~\cite{CHEN2019762,34_blp} and heuristics such as \emph{bipartite graph matching} based on optimal vertex assignments~\cite{27_Riesen} have been proposed, many of which support the graph edit distance in its full generality~\cite{34_blp,27_Riesen}. Several of these yield lower and upper bounds on the graph edit distance as a byproduct, which have just recently been compared systematically~\cite{29_ged_heuristics}.
However, these lower bounds for the general graph edit distance are not yet widely used for similarity search in graph databases.
For the methods based on optimal vertex assignments, it has only recently been shown how to derive a distance termed \textsc{Branch} that is guaranteed to be a lower bound and proven to satisfy the triangle inequality~\cite{92_improved}. \textsc{Branch} has been shown to provide an excellent trade-off between tightness and running time~\cite{29_ged_heuristics}.

We propose a filter-verification framework for similarity search, which supports the general graph edit distance with arbitrary metric edit costs and is hence suitable for graphs with any attributes comparable with a distance measure. We employ \textsc{Branch} as an initial filter accelerated by metric indexing. In the next stages, we derive upper bounds from the optimal assignment and improve them via local search to reduce the candidate set further, before performing verification by exact computation of the graph edit distance.
We experimentally evaluate our approach on graphs with attributes and categorical labels showing the effectivity of the filter pipeline. The results show that our approach allows scalable similarity search in attributed graphs with non-uniform edit costs. For the special case of uniform edit costs, where competing methods are available, our approach is shown to outperform the state of the art.

\section{Related Work}
\label{sec:relatedwork}
We discuss approaches for similarity search regarding the graph edit distance and methods for its pairwise exact or approximate computation.

\subsection{Similarity Search in Graph Databases}
\label{subsec:simsearchInGDB}

Methods for similarity search in graph databases can be divided into two categories, depending on whether they compare overlapping or non-overlapping substructures.
The methods \textit{k-AT}~\cite{12_efficient}, \textit{CStar}~\cite{2_ComparingStars}, \textit{Segos}~\cite{31_SEGOS} and \textit{GSim}~\cite{32_GSimJoin} belong to the first category.
These techniques are inspired by the \textit{$q$-grams} used in the computation of the string edit distance. Either $q$-grams based on trees~\cite{12_efficient,2_ComparingStars,31_SEGOS} or paths~\cite{32_GSimJoin} are used.
The methods \textit{Pars}~\cite{3_partitionSim}, \textit{MLIndex}~\cite{5_simMultiIndex}, and \textit{Inves}~\cite{21_inves2019} partition the graphs into non-overlapping substructures. They essentially obtain lower bounds based on the observation, that if $x$ non-overlapping substructures of a database graph are not contained in the query graph, the graph edit distance is at least $x$.
\textit{Pars} uses a dynamic partitioning approach to achieve this, while \textit{MLIndex} uses a multi-layered index to manage multiple partitions for each graph.
\textit{Inves} is a method used to verify whether the graph edit distance of two graphs is below a specified threshold by first trying to generate enough mismatching non-overlapping substructures.
\textit{Mixed}~\cite{ZhengZLWZ15} combines the idea of $q$-grams and graph partitioning. 
These methods only allow uniform edit costs and are therefore not suitable for graphs with continuous attributes. 

The concept of a \emph{median graph} of a set of graphs regarding the graph edit distance has been studied extensively, see~\cite{infosys2021a} and references therein. An application of median graphs is their use as routing objects in hierarchical index structures~\cite{103a_gedmetricindex,infosys2021a}. However, we are not aware of any concrete realization using this concept in a setting comparable to ours.

\subsection{Pairwise Computation of the Graph Edit Distance}
\label{subsec:pairwiseGEDComputation}
For computing the exact graph edit distance, both general-purpose algorithms~\cite{34_blp} as well as approaches tailored to the verification step in graph databases have been proposed~\cite{Chang2020}, which are usually based on depth- or breadth-first search~\cite{Gouda2016,Chang2020}, or integer linear programming~\cite{34_blp}.
As the exact computation of the graph edit distance is not feasible for larger graphs, many heuristics have been proposed, e.g., ~\cite{1_gedLinear,27_Riesen,29_ged_heuristics,35_beamS,92_improved,DBLP:conf/sisap/GoudaAC16}.
The properties of the dissimilarities obtained from these are in general not well investigated.
For heuristics based on optimal vertex assignment~\cite{27_Riesen}, which are widely used in practice~\cite{stauffer:2017aa}, a variant called \textsc{Branch} was recently studied thoroughly~\cite{92_improved}.
\textsc{Branch} is a lower bound on the graph edit distance, a pseudo-metric on graphs and supports arbitrary cost models (c.f., Section~\ref{sec:branch}).

\section{Preliminaries}
\label{sec:preliminaries}
We introduce the required basic concepts of graph theory and discuss database search with a focus on the metric space.

\subsection{Graph Theory}
\label{subsec:graphs}
A \emph{graph} $G = (V,E,\mu,\nu)$ consists of
a set of vertices $V(G)$,
a set of edges $E(G) \subseteq V(G)\times V(G)$ between vertices of $G$,
a labeling function for the vertices $\mu \colon V(G) \rightarrow L$, and  
a labeling function for the edges $\nu\colon  E(G) \rightarrow L$. 
We discuss only undirected graphs and denote an edge between $u$ and $v$ by $uv$.
The set of neighbors of a vertex $v \in V(G)$ is denoted by $N(v) = \{u\mid  uv\in E(G)\}$.
The set $L$ can be categorical labels or arbitrary attributes including real-valued vectors and complex objects such as strings.

A measure commonly used to describe the dissimilarity of two graphs is the \textit{graph edit distance}, which is the minimum cost for transforming one graph into the other using edit operations.
An \textit{edit operation} can be deleting or inserting an isolated vertex or an edge or relabeling any of the two.
An \textit{edit path} from graph $G_1$ to $G_2$ is a sequence of edit operations $(e_1, e_2, \dots)$ that transforms $G_1$ into~$G_2$.

\begin{definition}[Graph Edit Distance~\cite{27_Riesen}]
	Let $c$ be an edit cost function assigning non-negative costs to edit operations.
	The \emph{graph edit distance} between two graphs $G_1$ and $G_2$ is defined as
	$$d_{\text{ged}}(G_1,G_2) = \min\big\{ \textstyle\sum\nolimits_{i=1}^{k} c(e_i) \mid (e_1, \dots, e_k) \in \Upsilon(G_1,G_2) \big\},$$ where $\Upsilon(G_1,G_2)$ is the set of all possible edit paths from $G_1$ to $G_2$.
\end{definition}

\begin{table}[tb]\centering
	\caption{Notation for edit costs.}
	\label{tab:notedit}
	\begin{tabular}[c]{ll}
		\hline
		$c_v(u,v)$ & Cost of substituting vertex $u$ with vertex $v$ (adjusting the label/attributes)\\
		$c_v(u,\epsilon)$ & Cost of deleting the isolated vertex $u$\\
		$c_v(\epsilon,v)$ & Cost of inserting the isolated vertex $v$\\
		$c_e(uv,wx)$ & Cost of substituting edge $uv$ with edge $wx$ (adjusting the label/attributes)\\
		$c_e(uv,\epsilon)$ & Cost of deleting the edge $uv$\\
		$c_e(\epsilon,wx)$ & Cost of inserting the edge $wx$\\
		\hline
	\end{tabular}
\end{table}
The costs of the different edit operations can be chosen as required for the specific use case, see Table~\ref{tab:notedit} for our notation. If the edit costs are symmetric, non-negative, and strictly positive for each non-identical 
edit operation, the graph edit distance is a metric on graphs, treating graph isomorphism as identity~\cite{infosys2021a}. Note that this holds even if the edit costs do not satisfy the triangle inequality (and hence are no metric), because the graph edit distance uses the edit path with minimal cost.
In this work, we nonetheless assume that the edit costs respect the triangle inequality, i.e., we assume that the following inequalities hold:\footnote{For simplicity of notation, we have defined the costs on the vertices and edges instead of their labels. Hence, the sets $\mathcal{V}$ and $\mathcal{E}$ are all possible vertices and edges, respectively.}
\begin{alignat}{3}
c_v(u,w)&\leq c_v(u,v)+c_v(v,w) &\quad\forall (u,v,w)&\in \mathcal{V}^3 &\label{eq:triangle-1}\\
c_v(u,v)&\leq c_v(u,\epsilon)+c_v(\epsilon,v) &\quad\forall (u,v)&\in \mathcal{V}^2 &\label{eq:triangle-2}\\
c_e(uv,yz)&\leq c_e(uv,wx)+c_e(wx,yz) &\quad\forall (uv,wx,yz)&\in \mathcal{E}^3 &\label{eq:triangle-3}\\
c_e(uv,wx)&\leq c_e(uv,\epsilon)+c_e(\epsilon,wx) &\quad\forall (uv,wx)&\in \mathcal{E}^2 &\label{eq:triangle-4}
\end{alignat}
Equations (\ref{eq:triangle-1}), (\ref{eq:triangle-3}), and (\ref{eq:triangle-4}) can be enforced via pre-processing without changing the graph edit distance and can hence be assumed to hold w.l.o.g.~\cite{blumenthal:2020ac}. E.g., if we have $c(u,v){>}c(u,w){+}c(w,v)$, we can simply substitute $c(u,v)$ with $c(u,w){+}c(w,v)$, because a minimum cost edit path cannot contain $c(u,v)$. The only remaining constraint, Equation (\ref{eq:triangle-2}), is met (to the best of our knowledge) in all applications where the graph edit distance is used to address real-world problems~\cite{stauffer:2017aa}.
Computing the graph edit distance is \cNP-hard~\cite{3_partitionSim}, rendering exact computation possible for small graphs only.
There are several heuristics, many of which are based on solving an assignment problem.

\begin{definition}[Assignment Problem]
	\label{defAssProb}
	Let $A$ and $B$ be two sets with $|A|=|B|=n$ and $c\colon A \times B \to \mathbb{R}$ a cost function. 
	An \emph{assignment} between $A$ and $B$ is a bijection $f\colon A \to B$. The cost of an assignment $f$ is $c(f) = \sum_{a\in A} c(a,f(a))$.
	The \emph{assignment problem} is to find an assignment with minimum cost.
\end{definition}
For an assignment instance $(A,B,c)$, we denote the cost of an optimal assignment by $d^c_\mathrm{oa}(A,B)$.
The assignment problem can be solved in cubic running time using a suitable implementation of the Hungarian method~\cite{Burkard2012}.

\subsection{Searching in Databases}
\label{subsec:searchingInDB}

Databases provide means to store data to be able to retrieve, insert or change it efficiently.
In the context of data analysis, retrieval (search) is usually the crucial operation on databases,
because it will be performed much more often than updates.
We focus on two important types of similarity queries when searching a database $\DB$, the first of which is the \textit{range query} for a radius $r$:

\begin{definition}[Range Query]
	A \emph{range query} $\range{q}{r}$, with query object $q$ and range $r$, returns all objects in the database with a distance to the query object not exceeding the range, i.e.,
	$\range{q}{r}= \{o \in \DB \mid d(o,q)\leq r\}$.
\end{definition}
The second type of query considered here is the \textit{$k$-nearest neighbor query}.

\begin{definition}[$k$-Nearest Neighbor Query] 
	A $k$-nearest neighbor query \\(\kNNs query) $\kNN{q}{k}$ with query object $q$ and parameter $k$ returns the smallest set $\kNN{q}{k}\subseteq \DB$, so that
	$|\kNN{q}{k}|\geq k$ and
	
	$\forall o \in \kNN{q}{k}, \forall o^\prime\in \DB \setminus \kNN{q}{k}:\enskip d(o,q)< d(o^\prime,q).$
\end{definition}
In conjunction with range queries, it is preferable to return all the objects with a distance (including the query object, if part of the database),
that does not exceed the distance to the $k$th neighbor, which may be more than $k$ objects when tied.
That yields an equivalency of the results of \kNNs queries and range queries,
i.e., we have
$\range{q}{r}{=}\kNN{q}{|\range{q}{r}|}$
and $\kNN{q}{k}{=}\range{q}{r_k}$,
where $r_k$ is the maximum distance in $\kNN{q}{k}$.
Provided that the distance used is a metric, both types of queries can be accelerated using metric indices.
In our work, we use the \emph{vantage point tree} (\emph{vp-tree})~\cite{DBLP:conf/soda/Yianilos93} as a classical method and the more recent \emph{cover tree}~\cite{BeygelzimerKL06}, because they are available in the ELKI framework~\cite{DBLP:journals/corr/abs-1902-03616}, but others could also be used.
While the vp-tree is a height balanced binary tree dividing the data into \emph{near} and \emph{far} halves of the dataset based on the median distance from the vantage point,
the cover tree controls the expansion rate by reducing the maximum radius in each level of the tree, branching out if necessary into multiple branches.
In both trees, queries are performed top-down by traversing all paths that cannot be dismissed using the routing objects and employing the triangle inequality. 

\section{Efficient Filtering for the General Graph Edit Distance}
\label{sec:ourContribution}
\begin{figure}[tb]
	\centering
	\includegraphics[width=0.99\linewidth]{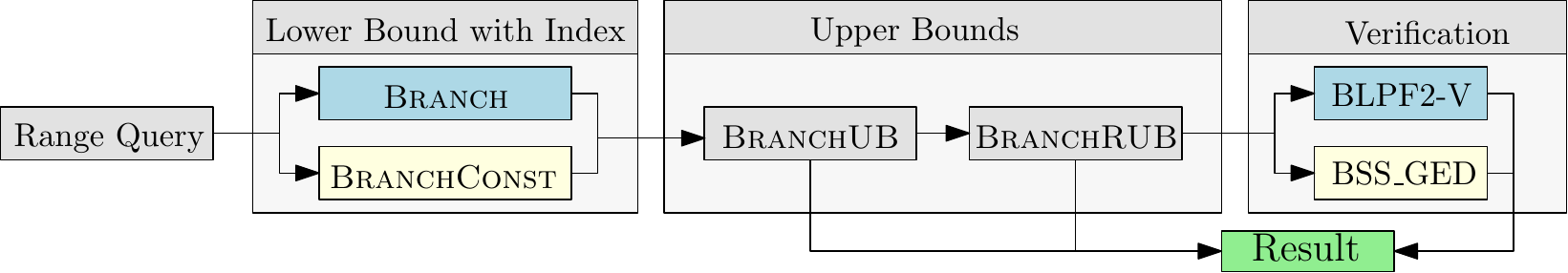}
	\caption{Overview of the filter pipeline. For general metric edit costs the blue modules are used; yellow modules are more efficient for uniform (edge) edit costs.} 
	\label{fig:filterpipeline}
\end{figure}
We propose a filter pipeline for range queries regarding the graph edit distance following a common paradigm for expensive distances, see e.g.~\cite{2_ComparingStars}. Here, lower bounds allow to filter out graphs that do not satisfy the query predicate. For the remaining candidates, upper bounds are evaluated to add them immediately to the result set without exact distance computation. Finally, in the verification step, the exact distance is computed for the remaining candidates only.
Our approach starts with the optimal assignment based lower bound \textsc{Branch} accelerated by metric indexing. From the same optimal assignment, an upper bound is derived (\textsc{BranchUB}) and subsequently refined by local search (\textsc{BranchRUB}) before the remaining candidates are verified. The pipeline is illustrated in Figure \ref{fig:filterpipeline}, the individual steps are described in the following.

\subsection{Index-Accelerated Lower Bound Filtering}\label{sec:branch}
Several lower bounds on the graph edit distance have been proposed or can be derived from known heuristics, see~\cite{29_ged_heuristics}. One of the most effective lower bounds with an excellent trade-off between tightness and runtime is referred to as \textsc{Branch}. 

\begin{definition}[Branch Distance] \label{branch}
	For two graphs $G_1$ and $G_2$ the \emph{branch distance} is defined as $d_{\text{branch}}(G_1,G_2) =d^c_\mathrm{oa}(V(G_1)\cup \varepsilon_1,V(G_2)\cup \varepsilon_2),$ where $\varepsilon_i$ denotes a multiset of $\epsilon$ elements, so that $|V(G_{i})\cup \varepsilon_i| = |V(G_1)\cup V(G_2)|$ for $i \in\{1,2\}$, and
	$$c(u,v) = \begin{cases}
	0    & \text{ if } u=v=\epsilon \\
	c_v(u,v)+ d_e(u,v) & \text{ if } u \neq \epsilon \text{ and } v \neq \epsilon \\
	c_v(\epsilon,v) + \nicefrac{1}{2} \cdot \sum_{n\in N(v)} c_e(\epsilon,vn)  & \text{ if } u=\epsilon \text{ and } v \neq \epsilon\\
	c_v(u,\epsilon) + \nicefrac{1}{2} \cdot \sum_{n\in N(u)} c_e(un,\epsilon)  & \text{ if } u \neq \epsilon \text{ and } v=\epsilon\\
	\end{cases},$$
	with $d_e(u,v)= d^{c'}_\mathrm{oa}(N(u)\cup \varepsilon_u,N(v)\cup \varepsilon_v) \text{, where } c'(w,x)=\nicefrac{1}{2} \cdot c_e(uw,vx)$.
\end{definition}

\begin{remark}
	Note that, by using a customized version of the Hungarian algorithm, \textsc{Branch} can also be implemented in a slightly more efficient way, where only one dummy vertex $\epsilon$ is added to the vertex sets $V(G_1)$ and $V(G_2)$ (see~\cite{bougleux:2020aa} for details). In this paper, we use the classical implementation employed in~\cite{27_Riesen,92_improved}, which corresponds to the characterization provided in Definition~\ref{branch}.
\end{remark}

\textsc{Branch} has its origin in one of the most successful heuristics for the graph edit distance proposed by Riesen and Bunke~\cite{27_Riesen}. However, in contrast to the original approach, it is guaranteed to underestimate the graph edit distance by dividing all edge costs by two to avoid that the cost of a single edge edit operation is counted twice, once for each endpoint~\cite{92_improved}.
Since an instance of the assignment problem on the vertices of the two graphs has to be solved, and for each vertex pair an assignment on their edges, \textsc{Branch} can be computed in $O(n^2\Delta^3 + n^3)$ time for graphs with $n$ vertices and maximum degree $\Delta$.
In the case of uniform edge edit costs, $d_e$ can be computed by multiset intersection of edge labels and the running time reduces to $O(n^3)$. This special case is referred to as \textsc{BranchConst}~\cite{92_improved}.
It has been shown that, if the edit costs are metric, the branch distance is a pseudo-metric on graphs~\cite{92_improved}. This allows to accelerate computing the candidate set w.\,r.\,t.{} this lower bound by employing metric indexing.

\subsection{Upper Bound Filtering and Verification}
From the solution of the assignment problem of \textsc{Branch}, an upper bound can be obtained by deriving the corresponding edit path~\cite{27_Riesen}, denoted \textsc{BranchUB} here. By definition of the graph edit distance, the cost of every edit path is an upper bound of the graph edit distance.
Following~\cite{2_ComparingStars}, we refine the assignment by local search to gain a tighter upper bound. Starting with the assignment obtained for the lower bound, the mapping of two vertex pairs is iteratively swapped, and kept whenever it induces a cheaper edit path, until there is no improvement. 
We refer to the refined upper bound obtained from the \textsc{Branch} assignment as \textsc{BranchRUB}.

Eventually, the graphs that were neither filtered out by the lower bound nor approved by the upper bounds are verified by exact graph edit distance computation.
We use \textit{BSS\_GED}~\cite{CHEN2019762} for datasets with discrete labels and uniform costs and \textit{BLPF2-V} otherwise. The latter is based on the integer programming formulation F2 of~\cite{34_blp} with the additional constraint that the objective function does not exceed the threshold to allow for early termination.

\subsection{Nearest-Neighbor Queries} \label{sec:nnopt}
For \kNNs queries it is not possible to separate the different steps of the filter pipeline as clearly as shown in Figure~\ref{fig:filterpipeline}. We realize \kNNs queries using the \emph{optimal multi-step $k$-nearest neighbor search} algorithm~\cite{25_optimalknn}. The database graphs are scanned in ascending order according the lower bound \textsc{Branch} regarding the query graph. For each graph, the exact graph edit distance is computed and the $k$ graphs with the smallest exact graph edit distance are maintained. 
Once we have found at least $k$ objects with an exact distance smaller than the lower bound of all remaining objects, the search can be terminated. This is optimal in the sense that none of the exact distance computations could have been avoided~\cite{25_optimalknn}. 
Accessing the graphs ordered regarding the \textsc{Branch} lower bound can be achieved na\"ively by sorting all graphs, or by using suitable metric index structures.

\section{Experimental Evaluation}
\label{sec:evaluation}

In this section, we experimentally address the following research questions:
\begin{itemize}
	\item[Q1] What speed-up in range queries can be achieved when using metric indices compared to a linear scan of the database?
	\item[Q2] How effective are the individual lower and upper bounds in our pipeline?
	\item[Q3] Can the proposed filter pipeline compete with state-of-the-art methods for uniform edit costs?
	\item[Q4] What speed-up in \kNNs queries can be achieved when using metric indices?
	\item[Q5] Does the proposed filter pipeline scale to a very large dataset?
\end{itemize}

\subsection{Setup}
\label{subsec:setup}
As metric index we chose the \textit{vp-tree} as a classical method and the \textit{cover tree} as a state-of-the-art approach.
For both we used the implementation provided by ELKI~\cite{DBLP:journals/corr/abs-1902-03616} with a sample size of $5$ for the vp-tree and an expansion rate of $1.2$ for the cover tree.

For a comparison in databases containing graphs with categorical labels, we used 
\textit{MLIndex}~\cite{5_simMultiIndex} and \textit{GSim}~\cite{32_GSimJoin}, since the former is considered state-of-the-art, while the latter provided much better results in our experiments. For \textit{MLIndex} the number of partitions was set to $\textrm{threshold}+1$ and in \textit{GSim} all provided filters were used.
We used the implementations provided by the authors.
In addition, we used \textit{CStar}~\cite{2_ComparingStars}, which follows a filter-verification approach related to ours.
For verification we used \textit{BSS\_GED}~\cite{CHEN2019762} and \textit{BLPF2-V}~\cite{34_blp} with the Gurobi solver.

\begin{table}[tb]\centering
	\caption{Datasets and their statistics~\cite{Datasets}. Some datasets contain graphs with labeled or attributed vertices and edges, as can be seen in the last two columns.
	}
	\label{tab:datasets}
	\small
	\begin{tabular}{lrcccc}
		\hline
		Name & $|$Graphs$|$ & avg $|$Vertices$|$ & avg $|$Edges$|$ & Labels (V/E) & Attributes (V/E) \\
		\hline
		\textit{Cuneiform}& $267$&$21.27$&$44.80$& +/+ & +/+\\
		\textit{Fingerprint}&$2800$&$5.42$&$4.42$&  --/-- & +/+\\
		\textit{Letter-high} & $2250$&$4.67$ &$4.50$& --/-- & +/--\\
		\textit{Letter-low} &$2250$&$4.68$&$3.13$&  --/-- & +/--\\
		\textit{MUTAG}&	$188$&	$17.93$&	$19.79$ & +/+ & --/-- \\
		\textit{PTC\_FM}&	$349$ &	$14.11$&	$14.48$ & +/+ & --/-- \\
		\textit{QM9}&	$129433$ &	$18.03$&	$18.63$& --/-- & +/+ \\	
		\hline
	\end{tabular}
\end{table}
We conducted experiments on a wide range of real-world datasets with different characteristics, see Table~\ref{tab:datasets}.
The costs of inserting, deleting or relabeling a vertex or edge with a categorical label were set to $1$, which is equivalent to the fixed setting in \textit{MLIndex}, \textit{GSim}, and \textit{CStar}. 
For continuous attributes, the Euclidean distance was used to define the relabeling cost. For simplicity, we did not use domain-specific distances.
Continuous attributes were normalized to the range $[0,1]$ (separately for each dimension), to make distances roughly comparable between different datasets.

\begin{figure*}[tb]
	\centering
	\includegraphics[width=0.99\linewidth]{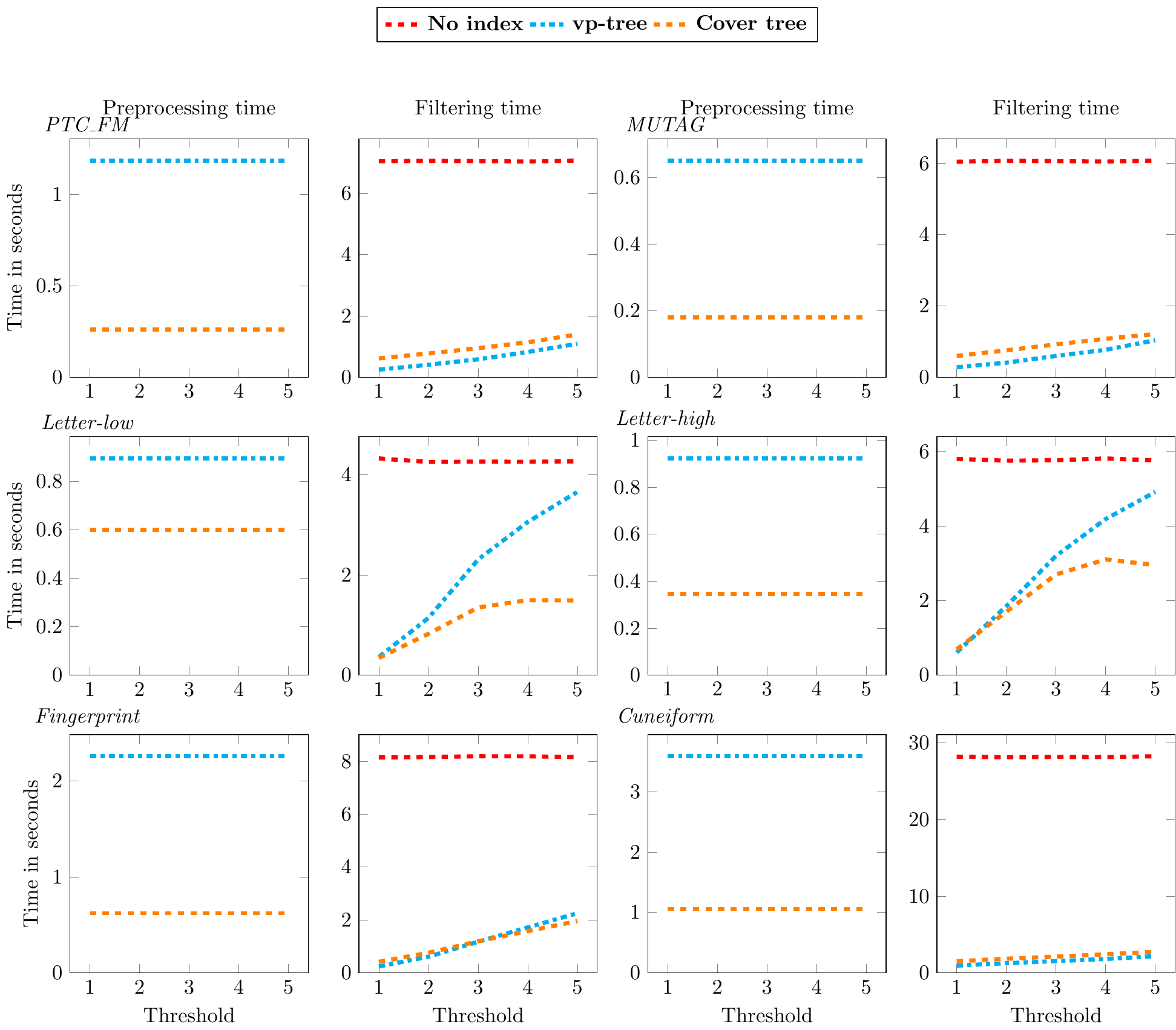}
	\caption{Runtime comparison for filtering $100$ range queries using \textsc{Branch} with thresholds $1$ to $5$ and  preprocessing time for constructing the index.}
	\label{fig:rangeruntimebranch}
\end{figure*}

\newpage
\subsection{Results}
\label{subsec:results}
We report on our findings regarding the above research questions.

\answer{Q1: Speed-up of range queries through metric indices.}
We first investigate how much of a speed-up can be achieved by using an index structure when filtering candidates for a range query by a lower bound.
We randomly sampled $100$ graphs from the dataset to be queries and then performed lower bound filtering without an index, using the cover tree, and the vp-tree.

Figure \ref{fig:rangeruntimebranch} shows the time needed for filtering $100$ range queries (each with thresholds $1$ to $5$) and additionally the preprocessing time  for index construction.
The runtime does not depend on the given threshold for a linear scan, but increases for the metric indices with the threshold, in particular for the \textit{Letter}-datasets.
It can be seen that, while on most datasets both index methods provide the same runtime benefit for filtering, the cover tree is much faster in preprocessing than the vp-tree. The runtime advantage on the \textit{Letter}-datasets is quite small for larger thresholds.
The runtime of the index structures directly corresponds to the number of \textsc{Branch} distance computations. Compared to the cover tree, the vp-tree requires many more distance computations in the preprocessing due to the chosen sample size. In general, the runtime corresponds to the number of candidates, which we investigate in the following.

\answer{Q2: Filter pipeline.}
\begin{figure*}[tb]
	\centering
	\includegraphics[width=0.75\linewidth]{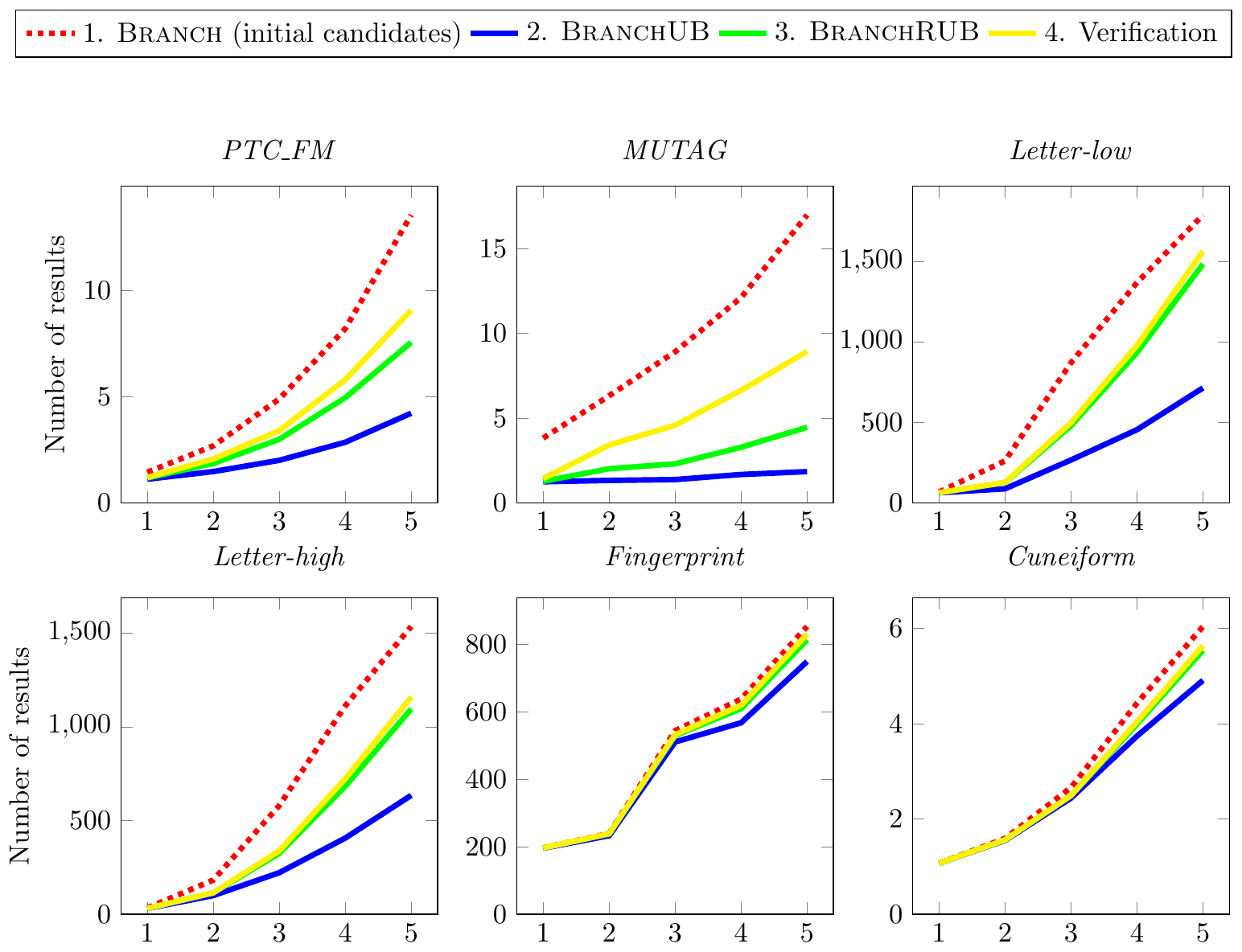}
	\caption{Average number of initial candidates (dashed) and hits identified in the different stages of the filter pipeline for each threshold.} \label{fig:candidatesbranch}
\end{figure*}
In this experiment we investigate how the candidate and result set are updated during filtering. 
Figure \ref{fig:candidatesbranch} shows the average number of candidates for \textsc{Branch} and the number of results after each step for $100$ range queries. When comparing the size of the candidate sets with the results of the previous experiment, it can be seen, that the runtime for filtering highly depends on the number of candidates.
For some datasets almost all candidates remaining after the upper bound filtering are not results. This indicates that improvement is possible with tighter lower bounds.
In general, \textsc{BranchRUB} manages to report almost all results, except in dataset \textit{MUTAG}.

\answer{Q3: Comparison with state-of-the-art methods.}
\begin{figure}[tb]
	\centering
	\includegraphics[width=0.99\linewidth]{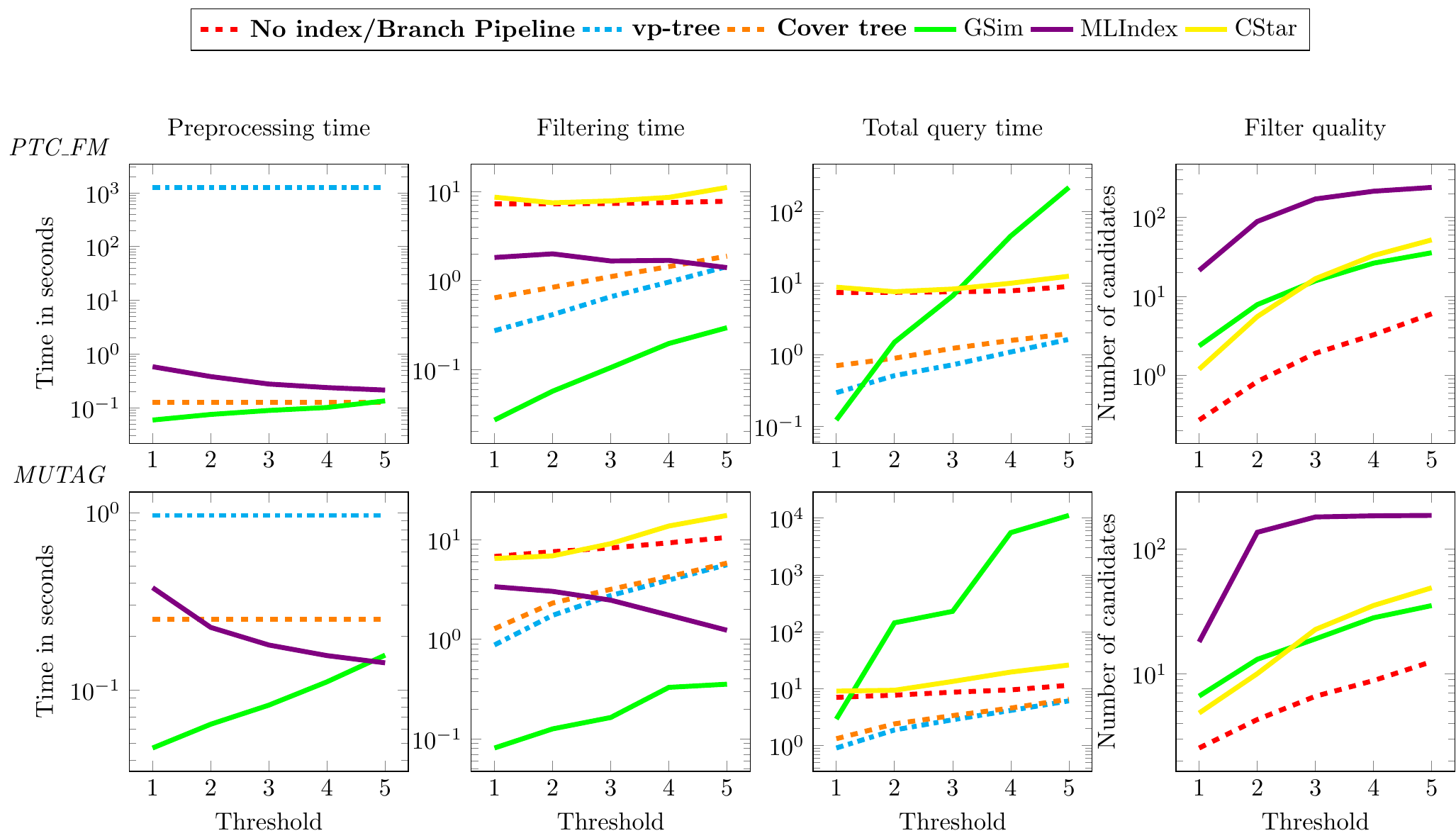}
	\caption{Runtime for answering $100$ range queries and average number of candidates remaining after applying all filters for thresholds $1$ to $5$. 
		Our approaches are shown with dashed lines and are marked bold in the legend. 
		For \textit{MLIndex} no verification time is given, since it did not finish within the time limit of $2$ days.}
	\label{fig:comparison}
\end{figure}
Many methods for similarity search in graph databases limited to uniform edit costs have been proposed. We compare to \textit{MLIndex}~\cite{5_simMultiIndex}, \textit{CStar}~\cite{2_ComparingStars} and \textit{GSim}~\cite{32_GSimJoin}. We used \textit{BSS\_GED}~\cite{CHEN2019762} for a fast verification in our filter pipeline, as well as in \textit{CStar}. The implementations of \textit{MLIndex} and \textit{GSim} contain their own verification algorithm. 
Figure~\ref{fig:comparison} shows the runtime for preprocessing and filtering as well as the total query time including filtering and verification for $100$ range queries. The average number of candidates remaining after application of all filters in the different methods is also shown. Only these need to be verified by exact graph edit distance computation. For \textsc{Branch} only one line is shown, since linear scan, \textit{vp-tree} and the \textit{cover tree} variant apply the same filters and generate the same candidates.

\textit{MLIndex} produces the largest candidate set, and did not finish the verification process in the time limit.
It can be seen that, while \textit{GSim} is quite fast in preprocessing and filtering, the verification step takes a long time. This is due to a combination of a slower verification algorithm and a higher number of candidates that have to be verified. The results indicate that, even when using \textit{BSS\_GED} for verification, the approach would not be competitive with the \textit{cover tree} due to the high number of candidates.
\textit{CStar} needs much more time for filtering and cannot filter out as many candidates leading also to a higher verification time.
Interestingly, the time for verification does not increase proportionally to the number of candidates, which might indicate, that the verification algorithm needs more time to verify certain \emph{difficult} graphs. 

\answer{Q4: Speed-up of \kNNs queries through metric indices.}
\begin{figure*}[tb]
	\centering
	\includegraphics[width=0.75\linewidth]{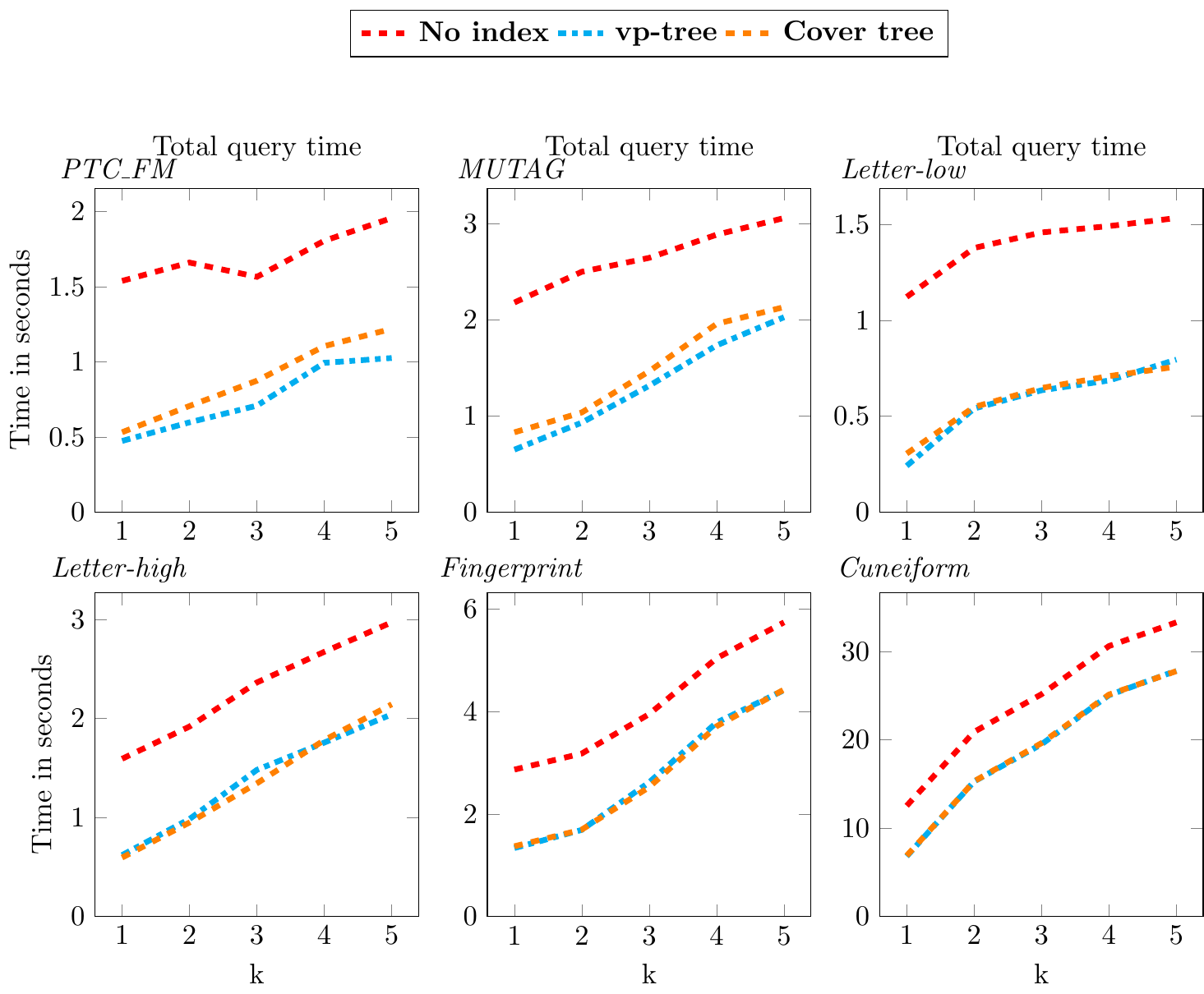}
	\caption{Runtime comparison for answering $20$ \kNNs queries using \textsc{Branch} and $k\in\{1,\dots,5\}$. 
	}
	\label{fig:knnruntimebranch}
\end{figure*}
We investigate how much of a speed-up can be achieved by using an index structure compared to not using one, when answering \kNNs queries using the optimal multi-step $k$-nearest neighbor search, cf.~Section~\ref{sec:nnopt}.
We randomly sampled $20$ graphs from the dataset to be queries and then used the cover tree as well as the vp-tree as the underlying metric index to compare them.
Figure~\ref{fig:knnruntimebranch} shows the time needed for answering $20$ \kNNs queries, each with $k\in\{1,\dots,5\}$ (excluding preprocessing).
Since in the optimal multi-step $k$-nearest neighbor search, the candidates have to be verified during search, before further candidates are explored, the runtime also includes the time needed for verification.
It can be seen that, again both index structures provide the same runtime benefit. Taking into account the preprocessing time however, the cover tree has a clear advantage over the vp-tree.

\answer{Q5: Similarity search in a large dataset.}
\begin{figure}[tb]
	\centering
	\includegraphics[width=0.5\linewidth]{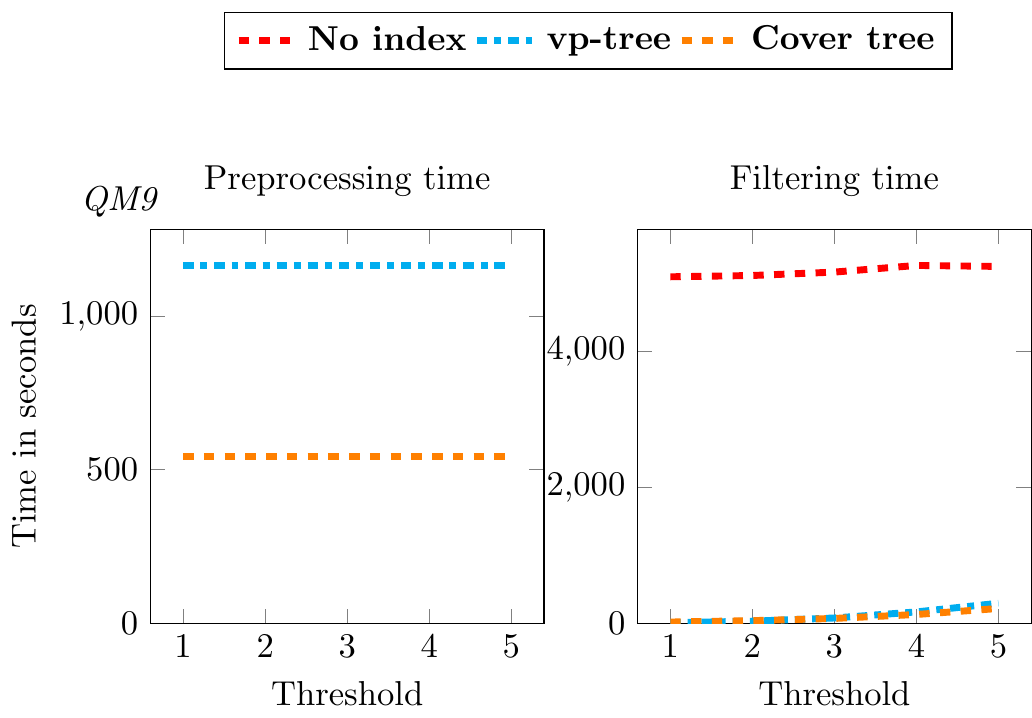}
	\caption{Runtime comparison for preprocessing and filtering $100$ range queries in the dataset \textit{QM9} using \textsc{Branch} and thresholds $1$ to $5$.}
	\label{fig:qm9rangeruntimebranch}
\end{figure}
We investigate the scalability of our approach on the dataset \textit{QM9} with $129\,433$ graphs with attributed vertices and edges.
The results shown in Figure~\ref{fig:qm9rangeruntimebranch} confirm the high preprocessing time of the vp-tree compared to the cover tree.
Both index methods achieve a significant advantage over a linear scan in filtering by reducing the running time by several orders of magnitude depending on the selectivity of the query.

\section{Conclusions}
\label{sec:conclusions}
We have shown that the recently studied lower and upper bounds on the graph edit distance can be employed to realize scalable graph similarity search in a filter-verification framework accelerated by metric indexing. Our approach supports attributed graphs without restrictions of edit costs.
For the extensively studied special case of graphs with discrete labels and uniform edit costs, our approach was shown experimentally to outperform the state-of-the-art methods.

There are several directions of future work to improve the filter-verification pipeline further. 
Our tightest upper bound was obtained via local search using a straightforward approach. More sophisticated techniques have been proposed recently~\cite{DBLP:journals/prl/BoriaBBB20} and can be incorporated to reduce verification. For the verification step, tailored methods that benefit from the already obtained assignment or the upper and lower bound can be developed.
A well-known phenomenon of metric trees is that their effectivity decreases with increasing intrinsic dimensionality of the data/distance.
Therefore, a suitable lower bound should not only be efficiently computed and tight, but ideally also have a low intrinsic dimensionality. Studying this property for the available lower bounds remains future work.
Finally, recent advances in median graph computation~\cite{infosys2021a} suggest to compute routing objects instead of using database graphs. An experimental comparison to such orthogonal approaches remains future work.

\setcitestyle{numbers}
\bibliography{lit}

\begin{thebibliography}{31}
\providecommand{\natexlab}[1]{#1}
\providecommand{\url}[1]{\texttt{#1}}
\expandafter\ifx\csname urlstyle\endcsname\relax
  \providecommand{\doi}[1]{doi: #1}\else
  \providecommand{\doi}{doi: \begingroup \urlstyle{rm}\Url}\fi

\bibitem[Beygelzimer et~al.(2006)Beygelzimer, Kakade, and
  Langford]{BeygelzimerKL06}
A.~Beygelzimer, S.~M. Kakade, and J.~Langford.
\newblock Cover trees for nearest neighbor.
\newblock In \emph{Int. Conf. Machine Learning, ICML}, volume 148, pages
  97--104, 2006.

\bibitem[Blumenthal and Gamper(2018)]{92_improved}
D.~B. Blumenthal and J.~Gamper.
\newblock Improved lower bounds for graph edit distance.
\newblock \emph{IEEE Transactions on Knowledge and Data Engineering},
  30\penalty0 (3):\penalty0 503--516, 2018.

\bibitem[Blumenthal and Gamper(2020)]{blumenthal:2020ac}
D.~B. Blumenthal and J.~Gamper.
\newblock On the exact computation of the graph edit distance.
\newblock \emph{Pattern Recognit. Lett.}, 134:\penalty0 46--57, 2020.

\bibitem[Blumenthal et~al.(2020)Blumenthal, Boria, Gamper, Bougleux, and
  Brun]{29_ged_heuristics}
D.~B. Blumenthal, N.~Boria, J.~Gamper, S.~Bougleux, and L.~Brun.
\newblock Comparing heuristics for graph edit distance computation.
\newblock \emph{{VLDB} J.}, 29\penalty0 (1):\penalty0 419--458, 2020.

\bibitem[Blumenthal et~al.(2021)Blumenthal, Boria, Bougleux, Brun, Gamper, and
  Gaüzère]{infosys2021a}
D.~B. Blumenthal, N.~Boria, S.~Bougleux, L.~Brun, J.~Gamper, and B.~Gaüzère.
\newblock Scalable generalized median graph estimation and its manifold use in
  bioinformatics, clustering, classification, and indexing.
\newblock \emph{Inf. Syst.}, 100:\penalty0 101766, 2021.

\bibitem[Boria et~al.(2020)Boria, Blumenthal, Bougleux, and
  Brun]{DBLP:journals/prl/BoriaBBB20}
N.~Boria, D.~B. Blumenthal, S.~Bougleux, and L.~Brun.
\newblock Improved local search for graph edit distance.
\newblock \emph{Pattern Recognit. Lett.}, 129:\penalty0 19--25, 2020.

\bibitem[Bougleux et~al.(2020)Bougleux, Gaüzère, Blumenthal, and
  Brun]{bougleux:2020aa}
S.~Bougleux, B.~Gaüzère, D.~B. Blumenthal, and L.~Brun.
\newblock Fast linear sum assignment with error-correction and no cost
  constraints.
\newblock \emph{Pattern Recognit. Lett.}, 134:\penalty0 37--45, 2020.

\bibitem[Burkard et~al.(2012)Burkard, Dell'Amico, and Martello]{Burkard2012}
R.~E. Burkard, M.~Dell'Amico, and S.~Martello.
\newblock \emph{Assignment Problems}.
\newblock {SIAM}, 2012.

\bibitem[Chang et~al.(2020)Chang, Feng, Lin, Qin, Zhang, and Ouyang]{Chang2020}
L.~Chang, X.~Feng, X.~Lin, L.~Qin, W.~Zhang, and D.~Ouyang.
\newblock Speeding up {GED} verification for graph similarity search.
\newblock In \emph{Int. Conf. Data Engineering, {ICDE}}, pages 793--804, 2020.

\bibitem[Chen et~al.(2019)Chen, Huo, Huan, and Vitter]{CHEN2019762}
X.~Chen, H.~Huo, J.~Huan, and J.~S. Vitter.
\newblock An efficient algorithm for graph edit distance computation.
\newblock \emph{Knowledge-Based Systems}, 163:\penalty0 762--775, 2019.
\newblock ISSN 0950-7051.

\bibitem[Gouda and Hassaan(2016)]{Gouda2016}
K.~Gouda and M.~Hassaan.
\newblock {CSI{\_}GED}: An efficient approach for graph edit similarity
  computation.
\newblock In \emph{Int. Conf. Data Engineering, {ICDE}}, pages 265--276, 2016.

\bibitem[Gouda et~al.(2016)Gouda, Arafa, and
  Calders]{DBLP:conf/sisap/GoudaAC16}
K.~Gouda, M.~Arafa, and T.~Calders.
\newblock {BFST{\_}ED}: {A} novel upper bound computation framework for the
  graph edit distance.
\newblock In \emph{{SISAP}}, pages 3--19, 2016.

\bibitem[Kim et~al.(2019)Kim, Choi, and Li]{21_inves2019}
J.~Kim, D.~Choi, and C.~Li.
\newblock Inves: Incremental partitioning-based verification for graph
  similarity search.
\newblock In \emph{Extending Database Technology, {EDBT}}, pages 229--240,
  2019.

\bibitem[Kriege et~al.(2019)Kriege, Giscard, Bause, and Wilson]{1_gedLinear}
N.~M. Kriege, P.~Giscard, F.~Bause, and R.~C. Wilson.
\newblock Computing optimal assignments in linear time for approximate graph
  matching.
\newblock In \emph{ICDM}, pages 349--358, 2019.

\bibitem[Lerouge et~al.(2017)Lerouge, Abu{-}Aisheh, Raveaux, H{\'{e}}roux, and
  Adam]{34_blp}
J.~Lerouge, Z.~Abu{-}Aisheh, R.~Raveaux, P.~H{\'{e}}roux, and S.~Adam.
\newblock New binary linear programming formulation to compute the graph edit
  distance.
\newblock \emph{Pattern Recognit.}, 72:\penalty0 254--265, 2017.

\bibitem[Liang and Zhao(2017)]{5_simMultiIndex}
Y.~Liang and P.~Zhao.
\newblock Similarity search in graph databases: {A} multi-layered indexing
  approach.
\newblock In \emph{Int. Conf. Data Engineering, {ICDE}}, pages 783--794, 2017.

\bibitem[Morris et~al.(2020)Morris, Kriege, Bause, Kersting, Mutzel, and
  Neumann]{Datasets}
C.~Morris, N.~M. Kriege, F.~Bause, K.~Kersting, P.~Mutzel, and M.~Neumann.
\newblock {TUD}ataset: A collection of benchmark datasets for learning with
  graphs.
\newblock In \emph{ICML 2020 Workshop on Graph Representation Learning and
  Beyond, GRL+}, 2020.

\bibitem[Neuhaus et~al.(2006)Neuhaus, Riesen, and Bunke]{35_beamS}
M.~Neuhaus, K.~Riesen, and H.~Bunke.
\newblock Fast suboptimal algorithms for the computation of graph edit
  distance.
\newblock In \emph{Structural, Syntactic, and Statistical Pattern Recognition},
  pages 163--172, 08 2006.

\bibitem[Qin et~al.(2020)Qin, Bai, and Sun]{QinBS20}
Z.~Qin, Y.~Bai, and Y.~Sun.
\newblock Ghashing: Semantic graph hashing for approximate similarity search in
  graph databases.
\newblock In \emph{{ACM} {SIGKDD}}, pages 2062--2072, 2020.

\bibitem[Riesen and Bunke(2009)]{27_Riesen}
K.~Riesen and H.~Bunke.
\newblock Approximate graph edit distance computation by means of bipartite
  graph matching.
\newblock \emph{Image Vision Comput.}, 27\penalty0 (7):\penalty0 950--959,
  2009.

\bibitem[Schubert and Zimek(2019)]{DBLP:journals/corr/abs-1902-03616}
E.~Schubert and A.~Zimek.
\newblock {ELKI:} {A} large open-source library for data analysis - {ELKI}
  release 0.7.5 "{Heidelberg}".
\newblock \emph{CoRR}, abs/1902.03616, 2019.

\bibitem[Seidl and Kriegel(1998)]{25_optimalknn}
T.~Seidl and H.~Kriegel.
\newblock Optimal multi-step k-nearest neighbor search.
\newblock In \emph{{SIGMOD} Int. Conf. Management of Data}, pages 154--165,
  1998.

\bibitem[Serratosa et~al.(2012)Serratosa, Cort{\'e}s, and
  Sol{\'e}-Ribalta]{103a_gedmetricindex}
F.~Serratosa, X.~Cort{\'e}s, and A.~Sol{\'e}-Ribalta.
\newblock Graph database retrieval based on metric-trees.
\newblock In \emph{SSPR}, pages 437--447, 2012.
\newblock ISBN 978-3-642-34166-3.

\bibitem[Stauffer et~al.(2017)Stauffer, Tschachtli, Fischer, and
  Riesen]{stauffer:2017aa}
M.~Stauffer, T.~Tschachtli, A.~Fischer, and K.~Riesen.
\newblock A survey on applications of bipartite graph edit distance.
\newblock In \emph{GbRPR}, pages 242--252, 2017.

\bibitem[Wang et~al.(2012{\natexlab{a}})Wang, Wang, Yang, and Yu]{12_efficient}
G.~Wang, B.~Wang, X.~Yang, and G.~Yu.
\newblock Efficiently indexing large sparse graphs for similarity search.
\newblock \emph{IEEE Trans. Knowl. Data Eng.}, 24\penalty0 (3):\penalty0
  440--451, 2012{\natexlab{a}}.

\bibitem[Wang et~al.(2012{\natexlab{b}})Wang, Ding, Tung, Ying, and
  Jin]{31_SEGOS}
X.~Wang, X.~Ding, A.~Tung, S.~Ying, and H.~Jin.
\newblock An efficient graph indexing method.
\newblock In \emph{Int. Conf. Data Engineering, ICDE}, 2012{\natexlab{b}}.

\bibitem[Yianilos(1993)]{DBLP:conf/soda/Yianilos93}
P.~N. Yianilos.
\newblock Data structures and algorithms for nearest neighbor search in general
  metric spaces.
\newblock In \emph{SODA}, pages 311--321, 1993.

\bibitem[Zeng et~al.(2009)Zeng, Tung, Wang, Feng, and Zhou]{2_ComparingStars}
Z.~Zeng, A.~K.~H. Tung, J.~Wang, J.~Feng, and L.~Zhou.
\newblock Comparing stars: On approximating graph edit distance.
\newblock \emph{Proc. {VLDB} Endow.}, 2\penalty0 (1):\penalty0 25--36, 2009.

\bibitem[Zhao et~al.(2012)Zhao, Xiao, Lin, and Wang]{32_GSimJoin}
X.~Zhao, C.~Xiao, X.~Lin, and W.~Wang.
\newblock Efficient graph similarity joins with edit distance constraints.
\newblock In \emph{Int. Conf. Data Engineering, ICDE}, 2012.

\bibitem[Zhao et~al.(2013)Zhao, Xiao, Lin, Liu, and Zhang]{3_partitionSim}
X.~Zhao, C.~Xiao, X.~Lin, Q.~Liu, and W.~Zhang.
\newblock A partition-based approach to structure similarity search.
\newblock \emph{Proc. {VLDB} Endow.}, 7\penalty0 (3):\penalty0 169--180, 2013.

\bibitem[Zheng et~al.(2015)Zheng, Zou, Lian, Wang, and Zhao]{ZhengZLWZ15}
W.~Zheng, L.~Zou, X.~Lian, D.~Wang, and D.~Zhao.
\newblock Efficient graph similarity search over large graph databases.
\newblock \emph{{IEEE} Trans. Knowl. Data Eng.}, 27\penalty0 (4):\penalty0
  964--978, 2015.

\end{thebibliography}

\end{document}